\documentclass[twoside,reqno]{mbook}
\usepackage{epsfig,cite}
\usepackage{amssymb,amsmath}
\usepackage{times}
\begin{document}

\sloppy \raggedbottom

\setcounter{page}{1}

\newpage
\setcounter{figure}{0}
\setcounter{equation}{0}
\setcounter{footnote}{0}
\setcounter{table}{0}
\setcounter{section}{0}



\title{Shell-model calculations
with modern nucleon-nucleon potentials}

\runningheads{Shell-model calculations 
with modern nucleon-nucleon potentials}{A.~Covello, 
L.~Coraggio, A~Gargano, N.~Itaco}

\begin{start}
\author{A.~Covello}{1}, \coauthor{L.~Coraggio}{1},
\coauthor{A.~Gargano}{1}, \coauthor{N.~Itaco}{1}

\address{Dipartimento di Scienze Fisiche, Universit\`a di Napoli Federico II,\\ and Istituto Nazionale di Fisica Nucleare, \\
Complesso Universitario di Monte S. Angelo, Via Cintia, I-80126 Napoli, Italy}
{1}

\begin{Abstract}
We report on a study of neutron-rich even-mass Te isotopes in terms of the shell model employing a realistic effective interaction derived from the CD-Bonn nucleon-nucleon potential. The strong short-range repulsion of the latter is renormalized by constructing a smooth low-momentum potential, $V_{\rm low-k}$, that is used directly as input for the calculation of the effective interaction.
In this paper, after a brief review of the theoretical framework, we present our results. Comparison shows that they are in very good agreement with the available experimental data.

\end{Abstract}
\end{start}

\section{Introduction}

During the past several years, shell-model calculations employing realistic effective interactions derived from modern nucleon-nucleon ($NN$) potentials have given an accurate description of the spectroscopic properties of a number of nuclei around closed shells. 
A main difficulty encountered in this kind of calculations is the strong short-range repulsion contained in the bare $NN$ potential $V_{NN}$, which prevents its direct use in the derivation of the shell-model effective interaction $V_{\rm eff}$.
As is well known, the traditional way  to overcome this difficulty is the Brueckner
$G$-matrix method. 

Recently, a new approach has been proposed \cite{Bogner01,Bogner02} which consists in deriving from $V_{NN}$ a renormalized low-momentum potential, $V_{\rm low-k}$, that preserves the physics of the original potential up to a certain cutoff momentum 
$\Lambda$.  This is a smooth potential which can be used directly to derive 
$V_{\rm eff}$. We have shown \cite{Bogner02,Covello02,Covello03} that this approach  provides an advantageous alternative to the use of the $G$ matrix. 

The main aim of this paper is to give a brief survey of the $V_{\rm low-k}$ approach
and its practical application in realistic shell-model calculations. Below we first give an outline of the derivation of $V_{\rm low-k}$ and $V_{\rm eff}$. We then report on some results of our current study of nuclei around doubly magic $^{132}$Sn, focusing attention on even-mass Te isotopes. The final Section provides a brief summary.

\section{Theoretical Framework}

The shell-model effective interaction $V_{\rm eff}$ is defined, as usual, in the following way. In principle, one should solve a nuclear many-body Schr\"odinger equation of the form 
\begin{equation}
H\Psi_i=E_i\Psi_i ,
\end{equation}
with $H=T+V_{NN}$, where $T$ denotes the kinetic energy. This full-space many-body problem is reduced to a smaller model-space problem of the form
\vspace{-.1cm}
\begin{equation}
PH_{\rm eff}P \Psi_i= P(H_{0}+V_{\rm eff})P \Psi_i=E_iP \Psi_i .
\end{equation}
\noindent Here $H_0=T+U$ is the unperturbed Hamiltonian, $U$ being an auxiliary potential introduced to define a convenient single-particle basis, and $P$ denotes the projection operator onto the chosen model space.

As pointed out in the Introduction, we ``smooth out" the strong repulsive core contained in the bare $NN$ potential $V_{NN}$ by constructing a low-momentum  potential
$V_{\rm low-k}$. This is achieved by integrating out the high-momentum modes of $V_{NN}$ down to a cutoff momentum  $\Lambda$. This integration is carried out with the requirement that the deuteron binding energy and phase shifts of $V_{NN}$ up to $\Lambda$ are preserved by $V_{\rm low-k}$. This requirement may be satisfied by the following $T$-matrix equivalence approach. We start from the half-on-shell $T$ matrix for $V_{NN}$ 
\begin{equation}
T(k',k,k^2) = V_{NN}(k',k) + \wp \int _0 ^{\infty} q^2 dq  V_{NN}(k',q)
\frac{1}{k^2-q^2} T(q,k,k^2 )\,,
\end{equation}
where $\wp$ denotes the principal value and  $k,~k'$, and $q$ stand for the relative momenta. 
The effective low-momentum $T$ matrix is then defined by

\begin{multline}
T_{\rm low-k }(p',p,p^2) = V_{\rm low-k }(p',p)   \\
+ \,\wp \int _0 ^{\Lambda} q^2 dq  
V_{\rm low-k }(p',q) \frac{1}{p^2-q^2} T_{\rm low-k} (q,p,p^2)\ ,  \label{eq:ham}
\end{multline}
where the intermediate state momentum $q$ is integrated from 0 to the momentum cutoff 
$\Lambda$ and $(p',p) \leq \Lambda$. 
The above $T$ matrices are required to satisfy the condition 
\begin{equation}
T(p',p,p^2)= T_{\rm low-k }(p',p,p^2) \, ; ~~ (p',p) \leq \Lambda \,.
\end{equation}

The above equations define the effective low-momentum interaction $V_{\rm low-k}$, and it has been shown \cite{Bogner02} that they are satisfied by the solution:
\begin{equation}
V_{\rm low-k} = \hat{Q} - \hat{Q'} \int \hat{Q} + \hat{Q'} \int \hat{Q} \int
\hat{Q} - \hat{Q'} \int \hat{Q} \int \hat{Q} \int \hat{Q} + ~...~~,
\end{equation}
which is the well known Kuo-Lee-Ratcliff (KLR) folded-diagram expansion \cite{KLR71,Kuo90}, originally designed for constructing  shell-model effective interactions.
In the above equation $\hat{Q}$ is an irreducible vertex function whose intermediate states are all beyond $\Lambda$ and $\hat{Q'}$ is obtained by removing from $\hat{Q}$ its terms first order in the interaction $V_{NN}$. In addition to the preservation of the half-on-shell $T$ matrix, which implies preservation of the phase shifts, this $V_{\rm low-k}$ preserves the deuteron binding energy, since eigenvalues are preserved by the KLR effective interaction. 
For any value of $\Lambda$, the low-momentum effective interaction of equation (6) can be calculated very accurately using iteration methods. Our calculation of $V_{\rm low-k}$ is performed by employing the iterative implementation of the Lee-Suzuki method \cite{Suzuki80} proposed in \cite{Andreozzi96}.

The $V_{\rm low-k}$ given by the $T$-matrix equivalence approach mentioned above is not Hermitian. Therefore, an additional transformation is needed to make it Hermitian. To this end, we resort to the Hermitization procedure suggested in \cite{Andreozzi96}, which makes use of the Cholesky decomposition of symmetric positive definite matrices.  

Once the $V_{\rm low-k}$ is obtained, we use it, plus the Coulomb force for protons, as input interaction  for the calculation of the matrix elements of $V_{\rm eff}$.
The latter is derived by employing a folded-diagram method, which was previously applied to many nuclei \cite{Covello01} using $G$-matrix interactions. Since $V_{\rm low-k}$ is already a smooth potential, it is no longer necessary to calculate the $G$ matrix. We therefore perform shell-model calculations following the same procedure as described, for instance, in \cite{Jiang92,Covello97}, except that the $G$ matrix used there is replaced by $V_{\rm low-k}$. More precisely, we first calculate the so-called 
$\hat{Q}$-box \cite{Kuo80} including diagrams up to second order in the two-body interaction. The shell-model effective interaction is then obtained by summing up the $\hat{Q}$-box folded-diagram series using the Lee-Suzuki iteration method \cite{Suzuki80}.

Clearly, the starting point of any realistic shell-model calculation is the free $NN$ potential. There are, however, several high-quality potentials, such as Nijmegen I and Nijmegen II \cite{Stoks94}, Argonne $V_{18}$ \cite{Wiringa95}, and CD-Bonn \cite{Machleidt01}, which fit equally well ($\chi^2$/datum $\approx 1$) the $NN$ scattering data up to the inelastic threshold. This means that their on-shell properties are essentially identical, namely they are phase-shift equivalent.   

In our shell-model calculations we have derived the effective interaction from the CD-Bonn  potential. This may raise the question of how much our results may depend on this choice of the $NN$ potential. We have verified that shell-model effective interactions derived from phase-shift equivalent $NN$ potentials through the 
$V_{\rm low-k}$ approach do not lead to significantly different results. Here, by way of illustration, we present the results obtained for the nucleus $^{134}$Te. This nucleus has only two valence protons and thus offers the opportunity to test directly the matrix elements of the various effective interactions. In Figure 1 we show, together with the experimental spectrum, the spectra obtained by using the CD-Bonn, NijmII, and Argonne $V_{18}$ potentials, all renormalized through the $V_{\rm low-k}$ procedure with a cutoff momentum $\Lambda$=2.2 fm$^{-1}$. This value of $\Lambda$ is in accord with the criterion given in \cite{Bogner02}. From Figure 1 we see 
that the calculated spectra are very similar, the differences between the level energies not exceeding 80 keV. It is also seen that the agreement with experiment is very good for all the three potentials.

\begin{figure}[t]
\centerline{\epsfig{file=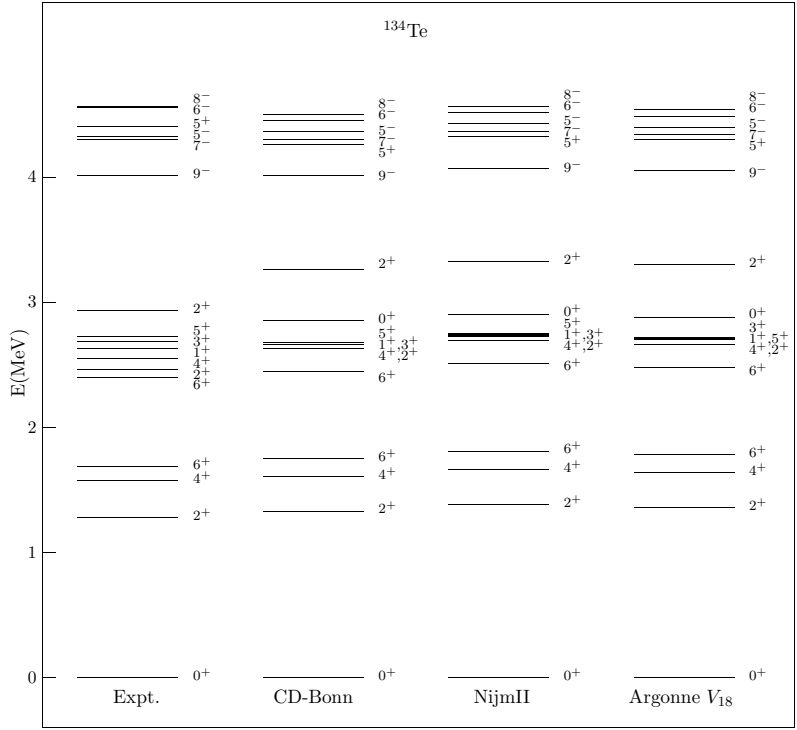,width=105mm}}
\caption{Spectrum of $^{134}$Te. Predictions by various $NN$ potentials are compared with experiment. \label{f1-ani}}
\end{figure}

\section{Calculations and Results}

We now present the results of our study of the even-mass Te isotopes
$^{132}$Te, $^{134}$Te, and  $^{136}$Te . The calculations have been performed using the OXBASH shell-model code \cite{OXBASH}.

We consider $^{132}$Sn as a closed core 
and let the valence protons and neutron holes occupy the five levels  $0g_{7/2}$, $1d_{5/2}$, $1d_{3/2}$, $2s_{1/2}$, and $0h_{11/2}$ of the 50-82 shell, while for the valence neutrons in $^{136}$Te the model space includes the six levels $0h_{9/2}$, $1f_{7/2}$, $1f_{5/2}$, $2p_{3/2}$, $2p_{1/2}$, and  $0i_{13/2}$ of the 82-126 shell.

As mentioned in the previous Section, the two-body matrix elements of the effective interaction are derived from the CD-Bonn $NN$ potential renormalized through the 
$V_{\rm low-k}$ procedure with a cutoff momentum $\Lambda=2.2$ fm$^{-1}$.  The computation of the diagrams included in the $\hat{Q}$-box is performed within the harmonic-oscillator basis using intermediate states composed of all possible hole states and particle states restricted  to the five shells above the Fermi surface. The oscillator parameter used is $\hbar \omega = 7.88$ MeV.

As regards the single-particle and single-hole energies, we have taken them from the experimental spectra of $^{133}$Sb, $^{133}$Sn, and $^{131}$Sn \cite{NNDC}. In the spectra of the two former nuclei, however, some single-particle levels are still missing. More precisely, this is the case of the proton $2s_{1/2}$ and neutron $0i_{13/2}$ levels, whose energies have been taken from \cite{Andreozzi97} and \cite{Coraggio02}, respectively, where it is discussed how they are determined. 
As regards the energy of the $h_{11/2}$ neutron hole state in $^{131}$Sn,  we have adopted the value of 0.100 MeV, which is in accord with the findings 
of \cite{Genevey00} and \cite{Fogelberg04}. This is somewhat smaller than that reported in \cite{NNDC}. 

\begin{figure}[h]
\centerline{\epsfig{file=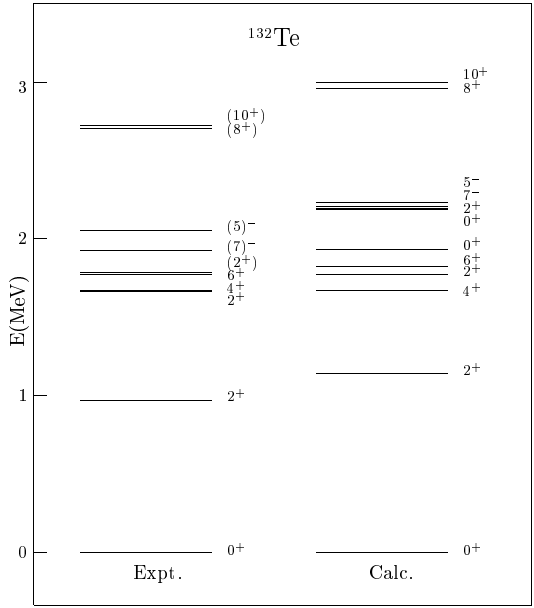,width=70mm}}
\caption{Experimental and calculated spectrum of $^{132}$Te. \label{f2-ani}}
\end{figure}

In Figure 2 we compare the experimental and calculated spectrum of $^{132}$Te. We see that while the theory reproduces all the observed levels, it also predicts 
the existence of two $0^+$ states around 2 MeV excitation energy. 

As regards the two-proton nucleus $^{134}$Te, the spectrum obtained by using the CD-Bonn potential has already been shown in Figure 1, where we see that it is in good agreement with the experimental one. Note that also in this case the theory predicts, around 2.5 MeV excitation energy, a $0^+$ state that has not been observed
so far. 

Let us now come to the two-proton, two-neutron nucleus $^{136}$Te.
From Figure 3 we see that also in this case the calculated spectrum reproduces nicely the experimental one.  

\begin{figure}[h]
\centerline{\epsfig{file=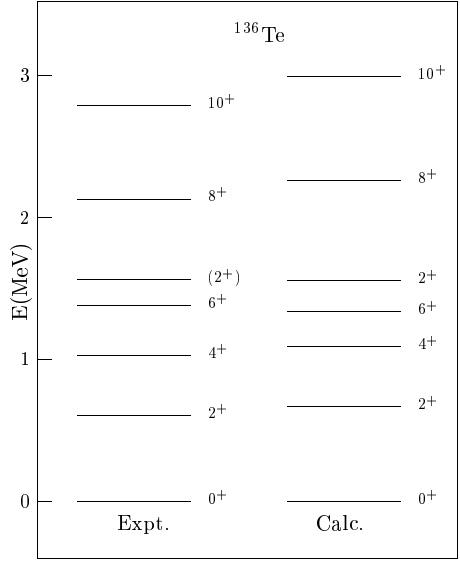,width=60mm}}
\caption{Experimental and calculated spectrum of $^{136}$Te. \label{f3-ani}}
\end{figure}

Recently \cite{Radford02}, the $B(E2;0^+ \rightarrow 2_1^+)$ values in  $^{132,134,136}$Te  have been measured using Coulomb excitation of neutron-rich radioactive ion beams. We have calculated these $B(E2)$'s with an effective proton charge of $1.55e$, according to our early study \cite{Andreozzi97} of $^{134}$Te. As for the effective neutron and neutron-hole charges, we have taken the values 0.70e and 0.78e, respectively, which have been adopted in our recent studies \cite{Coraggio02} and \cite{Genevey03}. The experimental \cite{Radford02} and theoretical $B(E2;0^+ \rightarrow 2_1^+)$ values are reported in Table 1, where we also compare our predictions with the other measured \cite{NNDC} $E2$ transition rates. 
\begin{table} [h]
\caption{\label{irreps} Experimental and calculated $B(E2)$ values (Wu) in
$^{132-136}$Te.}\smallskip
\begin{small}\centering
\begin{tabular*}{\textwidth}{@{\extracolsep{\fill}}cccc}
\hline  \noalign {\smallskip}
nucleus & $I_{i} \rightarrow I_{f}$ & Calc.  & Expt.  \\
\hline\noalign
{\smallskip}
${}^{132}$Te & $0^{+} \rightarrow 2^{+}$ & 38  & 43 $\pm$ 4  \\
& $6^{+} \rightarrow 4^{+}$  & 1.9 & 3.3 $\pm$ 0.2 \\
${}^{134}$Te & $0^{+} \rightarrow 2^{+}$   & 20 & $24 \pm 3$  \\
 & $4^{+} \rightarrow 2^{+}$  & 4.3 & $4.3 \pm 0.4$ \\
 & $6^{+} \rightarrow 4^{+}$  & 1.9 & $2.05 \pm 0.04$ \\
${}^{136}$Te & $0^{+} \rightarrow 2^{+}$ & 44 & $25 \pm 4$ \\
\\ \hline
\end{tabular*}
\end{small}
\end{table}

We see that our results are in very good agreement with the experimental values. In this connection, we should mention that a new measurement \cite{Baktash} of the $B(E2;0^+ \rightarrow 2_1^+)$ in $^{136}$Te
yields a value which is 50\% higher than that reported in Table 1, in excellent agreement with our prediction.

\section{Summary}

We have presented here the results of a shell-model study of nuclei in the close vicinity to $^{132}$Sn, focusing attention on the even-mass Te isotopes. The 
two-body effective interaction has been derived by means of a $\hat Q$-box 
folded-diagram method from the CD-Bonn $NN$ potential renormalized by use of the $V_{\rm low-k}$ approach. In this regard, we should emphasize that the $V_{\rm low-k}$'s extracted from various modern $NN$ potentials give very similar results in shell-model calculations, suggesting the realization of a nearly unique low-momentum $NN$ potential.
We have shown that our results are all in very good agreement with experiment. In particular, our calculations reproduce quite accurately the $B(E2;0^+ \rightarrow 2_1^+)$ values that have been recently measured through Coulomb excitation of radioactive ion beams.

\section*{Acknowledgments}

This work was supported in part by the Italian Ministero dell'Istruzione, dell'Universit\`a e della Ricerca (MIUR).


\begin{thebibliography}{99}

\bibitem{Bogner01} S. Bogner, T. T. S. Kuo, and L. Coraggio (2001) {\em Nucl. Phys. A} 
{\bf 684} 432c.
\bibitem{Bogner02}  S. Bogner, T. T. S. Kuo, L. Coraggio, A. Covello, and N. Itaco 
(2002) {\em Phys. Rev. C} {\bf 65} 051301(R). 
\bibitem{Covello02} A. Covello, L. Coraggio, A. Gargano, N. Itaco, and T. T. S. Kuo
(2002) in {\em Challenges of Nuclear Structure}, Proc. 7th Int. Spring Seminar on Nuclear Physics, (ed. A. Covello; World Scientific, Singapore) 139.
\bibitem{Covello03} A. Covello (2003), in {\em From Nuclei and their Constituents to Stars}, Proc. Int. School of Physics ``E. Fermi" Course CLIII, (ed. A. Molinari, L. Riccati, W. M. Alberico and M. Morando; IOS Press, Amsterdam) 79.
\bibitem{KLR71} T. T. S. Kuo, S. Y. Lee, and K. F. Ratcliff (1971), {\em Nucl. Phys. A}
{\bf 176} 65.
\bibitem{Kuo90} T. T. S. Kuo and E. Osnes (1990) {\em Lecture Notes in Physics}, Vol. ~ 364 (Springer-Verlag, Berlin).
\bibitem{Suzuki80} K. Suzuki and S. Y. Lee (1980) {\em Prog. Theor. Phys.} {\bf 64} 2091.
\bibitem{Andreozzi96} F. Andreozzi (1996) {\em Phys. Rev. C} {\bf 54} 684.
\bibitem{Covello01} A. Covello, L. Coraggio, A. Gargano, and N. Itaco (2001)
{\em Acta Phys. Pol. B} {\bf 32} 871, and references therein.
\bibitem{Jiang92} M. F. Jiang, R. Machleidt, D. B. Stout, and T. T. S. Kuo, 
(1992) {\em Phys. Rev. C} {\bf 46} 910.
\bibitem{Covello97} A. Covello, F. Andreozzi, L. Coraggio, A. Gargano, T. T. S. Kuo, and A. Porrino (1997) {\em Prog. Part. Nucl. Phys.}  {\bf 38} 165.
\bibitem{Kuo80} T. T. S. Kuo and E. M. Krenciglowa (1980) {\em Nucl. Phys. A} {\bf342} 454.
\bibitem{Stoks94} V. G. J. Stoks, R. A. M. Klomp, C. P. F. Terheggen, J. J. de Swart (1994), {\em Phys. Rev. C} {\bf 49} 2950.
\bibitem{Wiringa95} R. B.Wiringa, V. G. J. Stoks, and R. Schiavilla (1995) {\em Phys. Rev. C} {\bf 51} 38.
\bibitem{Machleidt01} R. Machleidt (2001) {\em Phys. Rev. C} {\bf 63} 024001.
\bibitem{OXBASH} B. A. Brown, A. Etchegoyen, and W. D. M. Rae, {\em The computer code OXBASH}, MSU-NSCL, Report No. 524.
\bibitem{Andreozzi97} F. Andreozzi, L. Coraggio, A. Covello, A. Gargano, T. T. S. Kuo, and  A. Porrino (1997) {\em Phys. Rev. C} {\bf 56} R16.
\bibitem{Coraggio02} L. Coraggio, A. Covello, A. Gargano, and  N. Itaco (2002) {\em Phys. Rev. C} {\bf 65} 051306(R).
\bibitem{NNDC} Data extracted usibg the NNDC On-line Data Service from the ENSDF database, file revised as of October 13, 2006.
\bibitem{Genevey00} J. Genevey, J. A. Pinston, H. Faust, C. Foin, S. Oberstedt, and B. Weiss (2000) {\em Eur. Phys. J. A} {\bf 7} 463.
\bibitem{Fogelberg04} B. Fogelberg {\em et al.} (2004) {\em Phys. Rev. C} {\bf 70} 034312.
\bibitem{Radford02} D. C. Radford {\em et al.} (2002) {\em Phys. Rev. Lett.} {\bf 88}
222501.
\bibitem{Genevey03}  J. Genevey, J. A. Pinston, H. R. Faust, R.Orlandi, A. Scherillo, G. S. Simpson, I. S. Tsekhanovic, A. Covello, A. Gargano, and W. Urban (2003)
{\em Phys. Rev. C} {\bf 67}, 054312.
\bibitem{Baktash} C. Baktash (private communication). 
\end{thebibliography}
\end{document}